# Disaster Information Reachability of Medium-Wave AM Broadcasting as a Wide-Area Disaster Information Infrastructure: Lessons from the 2024 Noto Peninsula Earthquake

Eiichi Shoji*
† *Advanced Materials Innovation & Monozukuri Lab, University of Fukui, 3-9-1 Bunkyo, Fukui 910-8507, JAPAN*
*Email:shoji@u-fukui.ac.jp, phone: +81-776-27-8076

**Abstract**

The 2024 Noto Peninsula Earthquake caused widespread power outages, communication failures, and road disruptions, and a tsunami warning was issued immediately after the earthquake, highlighting the importance of reliable disaster information transmission. The Noto Peninsula is surrounded by the sea on three sides and has limited transportation networks, making community isolation and securing reliable information transmission critical challenges. This study defines “Disaster Information Reachability” as the capacity to deliver necessary information to those who need it when they need it, and examines the role of medium-wave AM broadcasting from this perspective. In September 2024, the author conducted a mobile reception survey from Kanazawa to Suzu City and a fixed-point reception survey in Suzu City using commercially available radio receivers and passive radio receivers, including HOOPRA. Medium-wave stations in Niigata, Toyama, Akita, and Tokyo were clearly received, although the local NHK Kanazawa stations were affected by noise. The nationwide high-power NHK Radio 2 network, which was still operational at the time, was confirmed to provide a wide-area, multidirectional reception pathway through maritime propagation, demonstrating its critical significance as a redundant disaster information infrastructure. The fact that FM relay station damage was documented in subsequent policy materials, while medium-wave AM broadcasting received no explicit attention, suggests that its disaster information reachability remains insufficiently recognized. Medium-wave AM broadcasting, with its wide-area simultaneous broadcast capability, high-power coverage, and maritime propagation characteristics, functions effectively as a disaster information infrastructure in peninsular regions and warrants reassessment in preparation for large-scale disasters such as a Nankai Trough megaquake.

## 1. Introduction

At approximately 16:10 on January 1, 2024, the 2024 Noto Peninsula Earthquake (M7.6) struck with its epicenter in the Noto region of Ishikawa Prefecture [1]. The earthquake registered a maximum seismic intensity of 7 and developed into a large-scale disaster accompanied by tsunami generation, numerous casualties, and extensive housing damage [2]. Severe damage was also inflicted on social infrastructure, including roads, electricity, communications, and broadcasting, causing widespread power outages, communication failures, and transportation disruptions centered on the northern Noto Peninsula. Road damage was particularly serious: landslides, slope failures, road surface upheaval, and rockfalls occurred along major routes, including the Noto Satoyama Expressway (Fig. 1) and National Route 249, leaving numerous communities isolated. The Noto Peninsula is a typical peninsula region surrounded by sea on three sides (Fig. 2), and peninsula regions generally have limited road networks. Consequently, large-scale disasters tend to produce isolated communities through transportation network disruptions [3]. The Noto Peninsula region is also where aging has advanced far beyond the national average. In 2020, the aging rate reached approximately 44% in municipalities with a seismic intensity of upper-6 or above, approximately 52% in Suzu City, and approximately 46% in Wajima City, against a national average of approximately 28% [4]. The total population of the six affected municipalities also declined to approximately 61% of its 1985 level by 2020 [4], with depopulation and aging progressing simultaneously. These communities face persistent challenges in healthcare, transportation, and information transmission, even under normal conditions. The earthquake can be said to have made visible the inherent vulnerabilities of such peninsular regions. The Ministry of Land, Infrastructure, Transport, and Tourism has designated 23 regions nationwide as peninsula development promotion implementation areas under the Peninsula Development Act [5], including Tsugaru and Shimokita (Aomori Prefecture), Noto (Toyama and Ishikawa Prefectures), Kii (Mie, Nara, and Wakayama Prefectures), Sadamisaki (Ehime Prefecture), and Satsuma and Osumi (Kagoshima Prefecture), each facing the common challenges of limited road networks, aging populations, and depopulation. The Noto Peninsula Earthquake is therefore not merely a localized disaster, but an important case for considering the future of peninsular disaster prevention in Japan. Because peninsula regions sharing similar geographical conditions, such as the Kii and Sadamisaki regions (Fig. 2), are also expected to be heavily affected by the anticipated Nankai Trough megaquake, the knowledge gained from this earthquake constitutes an important lesson for future large-scale disasters. The Nankai Trough earthquake is a large-scale earthquake that has recurred at intervals of approximately 100–150 years; with approximately 80 years having elapsed since its most recent occurrence, its imminence is currently considered high [6]. Immediately after the earthquake struck, the author experienced it first in Fukui City, Fukui Prefecture, the neighboring prefecture. Watching NHK television at home, the author witnessed a broadcast in which an announcer urgently called for evacuation immediately after the tsunami warning was issued, following the emergency earthquake alert. The impassioned appeal — "Stop watching television and evacuate now" — subsequently became a topic of national discussion, but what came to the author's mind at that moment was the question: "How were the people of Oku-Noto able to listen to the radio during the earthquake?" In the Oku-Noto region, which faced the danger of a tsunami, up to approximately 44,160 households across Ishikawa Prefecture lost power immediately after the earthquake, and the affected area for mobile phone base stations reached a maximum of 70–82%. Roads were blocked at up to 93 locations, and up to 33 communities, comprising approximately 3,345 residents, were isolated. Housing damage reached 8,408 completely destroyed and 21,296 half-destroyed structures, and 455 sediment-related disasters also occurred, with compound damage simultaneously [4]. Telephone interviews conducted by the author subsequently confirmed that areas existed where power outages continued immediately after the disaster, leaving residents unable to watch television. The essential problem here is that evacuation information broadcast to the entire nation could not necessarily reach all those who needed it the most at the moment they needed it. People whose means of obtaining information are restricted during a disaster can be characterized as "information-vulnerable." The traditional concept of "disaster-vulnerable persons" has collectively referred to those with handicaps in their ability to perceive danger, obtain and transmit information, and take action [7]. The "information-vulnerable" as defined in this study shares elements with this concept while encompassing a broader perspective: it includes people who, despite being able-bodied under normal conditions, find it difficult to obtain information during disasters due to power outages, communication failures, geographical isolation, and constraints on their information device usage environment. As noted above, the aging rate in Suzu City stands at approximately 52% and across all municipalities with a seismic intensity of upper-6 or above at approximately 44%, making this perspective critically important in peninsula regions where aging and depopulation are advancing rapidly.

Regarding information transmission during disasters, in recent years, the proliferation of smartphones has driven advances in information systems using mobile networks, the Internet, social media, and various applications. However, in the 2024 Noto Peninsula Earthquake, the communication infrastructure itself sustained severe damage. According to the Ministry of Internal Affairs and Communications' "2024 White Paper on Information and Communications," up to approximately 7,800 fixed telephone lines and approximately 1,500 fixed internet lines were affected, and up to 839 mobile phone base stations were reported to have gone off-air [8]. The author's telephone interviews also yielded testimony that mobile phone signal bars decreased rapidly immediately after the disaster, after which calls became difficult to make, consistent with the white paper's records. During the recovery process, mobile base station vehicles, ship-based base stations, drone base stations, and satellite communication systems, including Starlink, were deployed; however, telephone interviews conducted by the author with local disaster prevention officials revealed that the satellite communication equipment needed to be installed outdoors with a clear line of sight in a cold, snowy environment, and that even slight positional shifts caused communication to become unstable. There were also cases in which attempts to use the equipment proved unsuccessful and were abandoned, demonstrating that the existence of advanced communication equipment and residents' ability to use it are not necessarily equivalent. Broadcasting infrastructure also sustained damage. For terrestrial television broadcasting, fuel replenishment became difficult owing to road disruptions, causing some relay stations to go off-air. Cable television networks, on which northern Noto Peninsula communities heavily depend, also experienced service interruptions due to power outages and transmission line breaks, with emergency restoration in Wajima and Suzu Cities extending to the end of March [4]. The trunk cables connecting the six municipalities of the northern Noto Peninsula (Wajima City, Suzu City, Noto Town, Anamizu Town, Nanao City, and Shika

Town) sustained widespread damage, rendering the regional information transmission network dysfunctional [4]. These are not problems unique to the Noto Peninsula, but challenges common to mountainous and peninsula regions across Japan, where depopulation is advancing. Meanwhile, the community FM station Radio Nanao temporarily went off-air but resumed broadcasting the following day, and efforts were also made to maintain medium-wave broadcasting continuity.

The author defines "Disaster Information Reachability" as the capacity to deliver necessary information to the people who need it at the moment they need it during a disaster. Disaster Information Reachability does not refer simply to communication performance or broadcast coverage area, but means "the capacity to deliver necessary information to the people who need it at the very moment they need it." This study focuses on medium-wave AM radio broadcasting (AM broadcasting) from the perspective of reachability. AM broadcasting possesses wide-area simultaneous broadcast capability through high-power transmitting stations, and receivers are inexpensive and operable at low power consumption. Medium-wave radio waves exhibit low propagation loss over seawater surfaces, enabling long-distance reach through maritime propagation [9, 10]. On the Noto Peninsula, which is surrounded by the sea on three sides, these maritime propagation characteristics make a substantial contribution to disaster information reachability. Furthermore, because information can reach the same area from multiple high-power transmitting stations, a high degree of redundancy is built into the disaster information transmission pathway. At the time of the 2024 Noto Peninsula Earthquake, a nationwide high-power medium-wave network, including NHK Radio 2, extensively covered the Japanese archipelago, possessing the latent capacity to deliver information from outside the disaster-affected area. The broadcast subsequently ended in late March 2026 [11], but the surveys and analyses in this study were based on conditions as they existed in September 2024, when the broadcast was still in operation. Following the full restoration of the Noto Satoyama Expressway, the author conducted on-site surveys in Suzu City using a commercially available radio receiver with signal strength display capability and passive radios (Hoop-shaped Radio: HOOPRA) [12–17] of 60 cm diameter for shoulder-carry use and 80 cm diameter for small fixed installation, developed by the author. As the HOOPRA also functions as an LC-resonant magnetically coupled receiving antenna, changes in reception sensitivity through non-contact magnetic coupling with a commercial radio were also confirmed. As described in detail below, clear reception was confirmed not only for stations within Ishikawa Prefecture, but also for distant medium-wave broadcasting stations in neighboring Fukui and Toyama Prefectures, as well as Niigata, Akita, and Tokyo. This study examines the role of maritime propagation characteristics of medium-wave AM broadcasting in disaster information reachability based on reception survey results from Suzu City. It aims to elucidate the reachability of AM broadcasting as an information infrastructure that "delivers to the end" during disasters, focusing on its wide-area simultaneous broadcast capability, redundancy, and maritime propagation characteristics. Reception conditions for AM and FM broadcasting during driving surveys along the Noto Satoyama Expressway and Suzu Road from Kanazawa to Suzu City were also reported.

## 2. Methods

### 2.1 Survey Area

Suzu City, located at the northern tip of the Noto Peninsula in Ishikawa Prefecture, was selected as the primary survey area (Fig. 1). Suzu City is a typical peninsula tip surrounded by the sea on three sides, with particularly limited transportation networks, and is among the areas that suffered serious damage in the 2024 Noto Peninsula Earthquake. The survey route followed the Noto Satoyama Expressway from Kanazawa City to Anamizu Town and then proceeded via the Suzu Road to Suzu City. The mobile reception survey was conducted after the expressway was fully restored. Within Suzu City, the area surrounding Suzu City Hall was designated the primary observation point (Fig. 1 and 3). This location has an open environment facing the Sea of Japan and is suitable for evaluating the maritime propagation of medium-wave broadcast waves arriving from various locations along the Sea of Japan coast. The survey was conducted on September 15, 2024, under the warning of heavy rain. Precipitation intensified after approximately 16:00, when the author departed [18], and additional surveys were not conducted in Wajima City for safety reasons.

### 2.2 Equipment

For the mobile reception survey, the AM/FM car radio of the Toyota Prius 20 series was used, representing a receiver close to actual disaster usage conditions. For the fixed-point reception survey, a commercially available radio receiver with signal-strength display capability (PL-368, TECSUN RADIO CO., LTD., Dongguan, China) was used for both AM and FM evaluations. FM reception uses only the built-in rod antenna, with no external antenna, to simulate realistic disaster conditions. In addition, passive radios (Hoop-shaped Radio: HOOPRA) developed by the author were used [12–17]. The HOOPRA operates solely on the energy of medium-wave broadcast waves without external power or batteries and can also improve the reception sensitivity of commercial radio receivers through non-contact magnetic coupling to the built-in ferrite-bar antenna. Units of 60 cm diameter (shoulder carry) and 80 cm diameter (small fixed installation) were used.

### 2.3 Survey Method

The survey comprised two stages: a mobile reception survey and a fixed-point reception survey. In the mobile survey, AM and FM reception conditions were continuously observed during travel from Kanazawa City to Suzu City, recording reception quality, noise, interruptions, and the need for frequency rescanning. In the fixed-point survey, reception conditions at Suzu City Hall were investigated using a commercial receiver and HOOPRA, recording frequency, broadcast content, transmitting station location, and reception quality. Reception quality was classified as "good reception" (broadcast content clearly audible), "receivable" (noise present but content identifiable), or "difficult reception" (content not identifiable); FM stations with unstable reception were classified as "unstable reception." To supplement the reception survey results, telephone interviews were conducted with individuals involved in disaster prevention and community management in the Oku-Noto region, covering conditions from immediately after the disaster through approximately two weeks afterward, including power outage conditions, communication environments, means of information acquisition, and the practical usability of satellite communication equipment.

### 2.4 Target Broadcasting Stations

The reception targets in this survey were medium-wave AM and FM broadcasting stations that were expected to be receivable from the area surrounding Suzu City Hall (Tables 1 and 2). In selecting the target stations, the geographical conditions of Suzu City — namely, its location as a peninsula tip surrounded by the Sea of Japan and Toyama Bay on three sides — were considered, and broadcasting stations from across the Sea of Japan coast were broadly included alongside stations within Ishikawa Prefecture.

As shown in Table 1, for medium-wave AM broadcasting, the survey targets included the NHK Kanazawa Broadcasting Station and the key and relay stations of Hokuriku Broadcasting (JOMR) within Ishikawa Prefecture, as well as stations in the neighboring prefectures of Toyama, Niigata, and Fukui, and distant high-power stations in Akita Prefecture and Tokyo. Specifically, the targets were: NHK Niigata Radio 1 (JOQK, 837 kHz, 10 kW, Akatsuka Broadcasting Station, Niigata City), NHK Niigata Radio 2 (JOQB, 1593 kHz, 10 kW, Akatsuka Broadcasting Station, Niigata City), Radio Niigata (JODR, 1116 kHz, 5 kW, Akatsuka Broadcasting Station, Niigata City), and NHK Niigata Radio 1 relay stations at Itoigawa Radio Relay Broadcasting Station, Itoigawa City (999 kHz, 0.1 kW), Takada Radio Relay Broadcasting Station, Joetsu City (792 kHz, 1 kW), and Kashiwazaki Radio Relay Broadcasting Station, Kashiwazaki City (981 kHz, 0.1 kW) in Niigata Prefecture; NHK Toyama Radio 1 (JOIG, 648 kHz, 5 kW, Toyota Broadcasting Station, Toyama City), NHK Toyama Radio 2 (JOIC, 1035 kHz, 1 kW, Toyama City), and Kitanihon Broadcasting (JOLR, 738 kHz, 5 kW, Toyota Broadcasting Station, Toyama City) in Toyama Prefecture; NHK Kanazawa Radio 1 (JOJK, 1224 kHz, 10 kW, Nonoichi Broadcasting Station), NHK Kanazawa Radio 2 (JOJB, 1386 kHz, 10 kW, Nonoichi Broadcasting Station), Hokuriku Broadcasting (JOMR, 1107 kHz, 5 kW, Nonoichi Broadcasting Station), and relay stations at Nanao

Radio Relay Broadcasting Station, Nanao City (NHK Radio 1: 540 kHz, 1 kW; NHK Radio 2: 1467 kHz, 0.1 kW; Hokuriku Broadcasting: 1107 kHz, 1 kW) and Wajima Radio Relay Broadcasting Station, Wajima City (NHK Radio 1: 1584 kHz, 0.1 kW; NHK Radio 2: 1359 kHz, 0.1 kW; Hokuriku Broadcasting: 1107 kHz, 0.1 kW) in Ishikawa Prefecture; NHK Fukui Radio 1 (JOFG, 927 kHz, 5 kW, Geba Broadcasting Station, Fukui City), NHK Fukui Radio 2 (JOFC, 1521 kHz, 1 kW, Geba Broadcasting Station, Fukui City), and Fukui Broadcasting (JOPR, 864 kHz, 5 kW, Sakai City) in Fukui Prefecture; NHK Akita Radio 2 (JOUB, 774 kHz, 500 kW, Akita Ogata Broadcasting Station) in Akita Prefecture; and NHK Tokyo Radio 1 (JOAK, 594 kHz, 300 kW, Shobu-Kuki Broadcasting Station, Kuki City) and NHK Tokyo Radio 2 (JOAB, 693 kHz, 500 kW, Shobu-Kuki Broadcasting Station, Kuki City). Note that the Hokuriku Broadcasting relay stations in Nanao City and Wajima City were suspended from operation at the time of the survey (September 15, 2024) because of the effects of the 2024 Noto Peninsula Earthquake.

As shown in Table 2, stations in Niigata, Toyama, Ishikawa, and Fukui Prefectures were targeted for FM broadcasting. Specifically, the targets were: NHK Niigata FM (JOQK-FM, 82.3 MHz, 1 kW, Niigata City), FM Radio Niigata (JOXU-FM, 77.5 MHz, 1 kW, Niigata City), and Niigata Broadcasting supplementary broadcast (92.7 MHz, 1 kW, Niigata City) in Niigata Prefecture; NHK Toyama FM (JOIG-FM, 81.5 MHz, 1 kW, Toyama City), Toyama FM Broadcasting (JOOU-FM, 82.7 MHz, 1 kW, Toyama City), and Kitanihon Broadcasting supplementary broadcast (90.2 MHz, 1 kW, Toyama City) in Toyama Prefecture; NHK Kanazawa FM (JOJK-FM, 82.2 MHz, 1 kW, Nonoichi City; 84.4 MHz, 0.1 kW, Nanao City; 83.2 MHz, 0.1 kW, Noto Town), FM Ishikawa (JOHV-FM, 80.5 MHz, 1 kW, Nonoichi City; 78.4 MHz, 0.1 kW, Nanao City; 81.9 MHz, 0.1 kW, Noto Town), and Hokuriku Broadcasting supplementary broadcast (94.0 MHz, 1 kW, Nonoichi City; 88.6 MHz, 0.1 kW, Nanao City; 76.7 MHz, 0.1 kW, Noto Town) in Ishikawa Prefecture; and NHK Fukui FM (JOFG-FM, 83.4 MHz, 1 kW, Fukui City), Fukui FM Broadcasting (JOLU-FM, 76.1 MHz, 1 kW, Fukui City), and Fukui Broadcasting supplementary broadcast (94.6 MHz, 1 kW, Fukui City) in Fukui Prefecture. Note that the FM relay stations located in Noto Town are referred to by name as "Suzu relay stations," but the actual broadcasting stations are located in Aza Akeno, Hosu-gun, Noto Town, Ishikawa Prefecture (in the vicinity of Takakura-yama), approximately 11 km from the Suzu City Hall survey point.

## 3. Results and Discussion

### 3.1 Information Acquisition Environment during the Noto Peninsula Earthquake

The 2024 Noto Peninsula Earthquake inflicted severe damage on communication infrastructure. Power outages and transmission line breaks together account for approximately 97% of mobile phone base station outages [19]. In the six municipalities of northern Noto Peninsula, NTT Docomo had a maximum of 260 base stations off-air, KDDI 248, and SoftBank 229, with outages continuing at high levels even during the critical 72-hour rescue period [19]. This reflected a structural problem: approximately 90% of mobile phone base stations nationwide have battery backup durations of less than 24 h, making early off-air status inevitable during prolonged power outages [19]. The areas affected by mobile phone services reached a maximum of 70% for NTT Docomo and 82% for KDDI [4], and landslides and road disruptions made fuel supply and maintenance difficult. The author's telephone interviews confirmed that mobile phone signal bars decreased immediately after the disaster, after which calls became impossible, and that congestion continued to hinder calls even after restoration.

During the recovery process, mobile base station vehicles, ship-based and drone base stations, portable generators, and satellite communication systems, including Starlink, were deployed, with several hundred Starlink units introduced at evacuation shelters and local government facilities. However, interviews with community center staff revealed that this equipment was not necessarily easy to use: it required outdoor installation with a clear line of sight in a cold, snowy environment, even slight positional shifts caused instability, and some attempts were abandoned outright, compounded by a shortage of personnel familiar with its operation. This demonstrates that the existence of communication equipment and residents' ability to use it are not necessarily equivalent.

Broadcasting infrastructure also sustained damage. For terrestrial television, road disruptions make fuel replenishment difficult for relay stations operating on emergency power. Seven relay stations in the Noto district were damaged, with an emergency power capacity of only one to five days [20]. The Wajima relay station avoided outages only because fuel was airlifted by Self-Defense Force helicopters six times, sustaining 20 days of continuous generator operation [20]; the Wajima-Machino relay station (649 households) was nonetheless off-air from January 2 to 24, and the Higashimonzen relay station (1,157 households) from January 2 to 4 [20]. NHK and four commercial television stations went off-air in parts of Wajima City, affecting approximately 2,130 households [8]. Cable television, on which northern Noto Peninsula communities heavily depend due to mountainous terrain limiting terrestrial reception, also suffered widespread trunk cable damage across the six municipalities, rendering the regional network dysfunctional, with emergency restoration in Wajima and Suzu Cities extending to the end of March [4, 8]. This demonstrates that the vulnerability of information infrastructure in peninsula regions is tied not merely to communications and broadcasting equipment in isolation but to the entirety of social infrastructure, including road and logistics networks — a structural vulnerability reproduced in the September 2024 Oku-Noto heavy rain disaster, when the Hegura relay station was off-air for nearly a month and the Wajima-Machino relay station for several days–in both cases because batteries were exhausted before commercial power was restored [20]. These problems are common in mountainous and peninsula regions across Japan and raise concern for the anticipated Nankai Trough megaquake [6]. Radio broadcasting was restored relatively quickly. The community FM station Radio Nanao went off-air briefly but resumed the following day, providing water supply and life support information through emergency interruption broadcasting under a disaster prevention agreement with Nanao City. Medium-wave broadcasting also maintains continuity through emergency power and rapid response. Nonetheless, the author's interviews revealed that in some areas, particularly the Sotouura region, radio reception had been unstable even under normal conditions, and a culture of regular radio use had not fully taken hold. Even so, in environments where power and communication fail, battery-powered radio and automobile radio maintain comparatively high availability. This is corroborated by the Great East Japan Earthquake, where AM radio received the highest disaster-time usefulness rating of any medium at 60.1% [8], with survivor testimony that "radio was the only means of obtaining information immediately after the disaster" [8]. The author's interviews confirmed similar conditions in the 2024 Noto Peninsula Earthquake, supporting the value of medium-wave AM broadcasting for disaster information reachability across multiple disaster cases. This gap between the existence of information transmission means and the actual delivery of information to residents corroborates the importance of the "Disaster Information Reachability" concept proposed in this study.

A maximum of approximately 40,000 households lost power at 16:11 on January 1, just one minute before the tsunami warning was issued at 16:12 [23]. Power transmission rates in the Oku-Noto region were only 17% in Wajima City, 12% in Suzu City, 3% in Anamizu Town, and 14% in Noto Town at the time of the disaster, confirming that the vast majority of households in tsunami-threatened coastal areas were already without power when the warning was issued [23]. Municipal-level outage figures show approximately 13,000 households in Wajima City, 7,800 in Suzu City, 8,900 in Noto Town, and 5,300 in Anamizu Town [24]. A post-disaster survey found television to be the most frequently cited medium for receiving tsunami warnings, at 69.6–72.5% [25]. However, this was a recall-based questionnaire that did not distinguish between information obtained while sheltering in place under power outage conditions, while evacuating, and after arriving at a shelter. As household television receivers become inoperable during power outages as an engineering truism, these figures do not directly indicate whether television was available during the critical initial phase at 16:12. The same survey's markedly higher television usefulness ratings among respondents who evacuated by automobile suggest that in-vehicle television contributed non-negligibly to the overall statistics. The survey sample also warrants scrutiny: respondents in the coastal areas of Wajima and Suzu Cities that actually sustained tsunami damage accounted for only 4.7% of the total [25], making it difficult to conclude that the most severely affected residents were adequately represented — a methodological limitation that must be considered when interpreting the results.

In contrast, a separate survey found that among respondents who experienced power outages, radio recorded the highest usage rate of any information device [26], consistent with the engineering reality that battery-powered and passive radio receivers remain operational during outages, and aligning with the reception survey results of the present study. The medium-wave AM broadcasting network itself was in operation at the time of the disaster, while the September 2024 survey subsequently confirmed that broadcast signals from multiple directions were receivable in Suzu City.

### 3.2 Radio Reception Conditions along the Road to Suzu City

The author traveled from Kanazawa City to Suzu City via the Noto Satoyama Expressway and Suzu Road after full restoration of the expressway, conducting a reception survey using car radio. This route is the primary arterial road on the Noto Peninsula and is a critical axis for relief activities and resident evacuation during disasters (Fig. 1). Both AM and FM broadcasts were well received near Kanazawa City, but a marked difference emerged, heading north. For FM broadcasting, station switching occurred frequently, with noise increasing to the point of difficulty in sections where the field strength decreased, repeatedly necessitating the car radio's auto-scan function. The reception quality deteriorated markedly in mountainous and hilly sections, and the number of receivable FM stations decreased near Suzu Road. Because FM broadcasting uses VHF band frequencies and relies primarily on line-of-sight propagation, it is highly susceptible to terrain shielding, with the reception environment becoming unstable when approaching the peninsula tip. This tendency also appeared in the fixed-point survey: the mountainous terrain between Suzu City and Nonoichi City (116 km away), where Ishikawa Prefecture's FM key stations are located, obstructs VHF line-of-sight propagation, yielding RSSI values of only 10–16 even within the same prefecture, demonstrating the fundamental vulnerability of FM broadcasting in peninsula regions. AM broadcasting, by contrast, was receivable with relative stability throughout the entire route. Station switching occurred but rarely required the frequent rescanning seen with FM, and broadcast content remained continuously audible even when the field strength decreased, with only a slight increase in reception noise (distinct from the noise generated by inverter devices and power equipment).

With AM broadcasting, audio information is maintained in many cases even at reduced reception levels, giving high practical information acquisition capability. Continuous reception was possible for NHK Radio 1, commercial AM broadcasts, and out-of-prefecture medium-wave stations throughout the drive.

During a disaster, the ability to continue obtaining information matters more than audio quality: FM offers high audio quality but loses information entirely when reception becomes impossible, whereas AM can continue delivering information despite some noise. From the perspective of disaster information reachability, AM broadcasting's wide-area propagation capability and reception continuity are therefore critically important characteristics in peninsular regions, such as the Noto Peninsula, consistent with the fixed-point survey results presented in the following section.

### 3.3 Medium-Wave Broadcast Reception Survey in Suzu City and Wide-Area Propagation Characteristics

Upon arrival at Suzu City, a fixed-point reception survey was conducted in the area surrounding Suzu City Hall using a commercially available radio receiver with signal strength display capability and passive radios (HOOPRA) developed by the author. Suzu City, located at the very tip of the Noto Peninsula, with Toyama Bay to the east and the Sea of Japan to the north and west, constitutes an extremely valuable observation point for evaluating maritime propagation characteristics. The initial survey using the commercial receiver confirmed reception of the NHK Kanazawa Broadcasting Station and commercial stations within Ishikawa Prefecture, although reception conditions were not necessarily optimal for local stations. For NHK Radio 1 in particular, reception was confirmed not only from Kanazawa but also from the Sea of Japan coastal stations, including Toyama and Niigata, and during nighttime hours, clear reception was confirmed from stations several hundred kilometers away, including Akita and Shobu-Kuki Broadcasting Stations (Kuki City). Similar results were obtained with HOOPRA: the 60 cm unit enabled reception of multiple distant stations, and the 80 cm unit achieved even higher performance, demonstrating sufficient medium-wave field strength at the survey location without any external power or batteries. At Suzu City Hall, magnetic coupling of the 80 cm HOOPRA unit to the PL-368 receiver substantially increased the received signal strength. The RSSI of NHK Kanazawa Radio 1 (1224 kHz) increased from 24 to 55, that of NHK Kanazawa Radio 2 (1386 kHz) increased from 15 to 46, and that of Fukui Broadcasting (864 kHz) increased from 24 to 50. Although this is a single example, it demonstrates that the reception enhancement provided by the HOOPRA antenna is effective, even in the Oku-Noto region. The 80 cm unit was selected because it provides the largest enhancement effect among the stationary types. Although lightweight flexible versions of the HOOPRA are also available, a rigid-frame unit was used to ensure a stable and reproducible coupling geometry during measurement. HOOPRA also enabled the reception of weak stations tending to be buried by noise with standard receivers, confirming the richness of the medium-wave environment. Stations confirmed as received included those in Ishikawa, Toyama, Niigata, Akita, and Tokyo, with even more distant stations confirmed depending on the time of day. Because many of these stations lie along the Sea of Japan coast, the majority of the propagation paths to Suzu City run over water. Medium-wave radio waves exhibit low propagation loss over seawater surfaces [9, 10] because seawater's high electrical conductivity ($\sigma$ = 5 S/m) [21] keeps ground wave attenuation small over longer distances than overland paths. Suzu City, surrounded by the sea on three sides, is therefore well-positioned to receive medium-wave broadcast waves from across the Sea of Japan coast.

The spatial distribution of the received medium-wave broadcasting stations around Suzu City is shown in Fig. 4, and the detailed results are presented in Table 1. Examining the relationship between straight-line distance and RSSI, no monotonic relationship was found; rather, the proportion of maritime path segments strongly influenced the reception quality. Toyama City (distance: 81 km, output: 5 kW) showed high RSSI values of 60–61, attributable to the combination of a short distance and a propagation path consisting primarily of maritime propagation over Toyama Bay. The Niigata City key station (distance: 145 km, output: 10 kW) yielded an RSSI of 50, where a higher output and a secured maritime path over the Sea of Japan compensated for the greater distance. Among the Niigata Prefecture relay stations, Itoigawa City (distance: 68 km, output: 0.1 kW) showed an RSSI of 34 and Joetsu City (distance: 91 km, output: 1 kW) showed an RSSI of 41. Despite Itoigawa's output being one-tenth of that of Joetsu, the small RSSI difference reflects Itoigawa's near-entirely maritime path (located near the mouth of the Hime River) combined with Joetsu's lower frequency of 792 kHz, which favors ground wave propagation despite its inland location and overland path segments. The NHK Nonoichi Broadcasting Station within Ishikawa Prefecture (distance: 116 km, output: 10 kW) showed a comparatively low RSSI of 24 because its propagation path included overland segments across the Noto Peninsula. The NHK Radio 1 relay station in Nanao City (distance: 53 km, 540 kHz, 1 kW) showed a high RSSI of 46, reflecting the small ground-wave attenuation achievable at this lower frequency, even at a modest output.

Of particular note is the comparison between NHK Akita Radio 2 (JOUB, 774 kHz, 500 kW, RSSI 40) at the Akita Ogata Broadcasting Station and NHK Tokyo Radio 1 and Radio 2 (RSSI 37 and 35, respectively) at the Shobu-Kuki Broadcasting Station (Kuki City). Despite being at the greatest distance from Suzu City (375 km), Akita Ogata yielded good reception with low noise, whereas Shobu-Kuki, 117 km closer at 258 km, showed lower RSSI values of 35–37 with considerable noise. The propagation paths of these distant high-power stations are shown in Fig. 5. This reflects the differing nature of the two propagation paths: Akita Ogata's path to Suzu City is almost entirely maritime over the Sea of Japan, maximizing the benefit of low-loss seawater propagation, whereas Shobu-Kuki's path passes through the Kanto Plain and mountainous terrain along the Niigata Prefecture border before reaching Toyama Bay, subjecting it to substantial overland propagation loss and scattering. These results clearly demonstrate that medium-wave reception quality depends not on the simple distance from the broadcasting station, but on the proportion of maritime segments in the propagation path. At a peninsula tip, such as Suzu City, radio waves from stations with secured maritime paths can deliver higher reception quality even at great distances than radio waves from stations routed over land at shorter distances.

An FM broadcast reception survey was conducted using the same methodology (Table 2). Because FM broadcasting relies on VHF-band line-of-sight propagation, it exhibits fundamentally different characteristics. The Noto Town relay station (distance: 11 km, output: 0.1 kW) showed outstandingly high RSSI values of 34–37 with good reception, demonstrating that the FM reception quality is strongly influenced by the presence of a nearby relay station. All other stations showed RSSI values of only 9–17, with noise generally dominant and audible reception extremely difficult at an RSSI of 20 or below. Toyama City (81 km) and Niigata City (142 km) both showed RSSI values of 14–17, despite a distance ratio of approximately 1.75, demonstrating that FM propagation cannot be explained by simple distance attenuation and depends heavily on terrain shielding. The mountainous terrain between Toyama and Suzu City contrasts with the comparatively less shielded path from Niigata, compensating for the distance difference. Fukui City (177 km) at RSSI 9–17 likely reflects similarly favorable sea-of-Japan-side geographical conditions. A comparison of the two tables — AM's RSSI 35–40 for stations 258–375 km away versus FM's RSSI 9–17 for all but the nearby relay station — clearly demonstrates that AM broadcasting achieves stable reception from high-power stations hundreds of kilometers away through maritime propagation, whereas FM did not achieve stable reception in this survey without a nearby relay station, concisely illustrating the difference in characteristics between the two as information infrastructure in peninsular regions.

In practice, Suzu City could simultaneously receive broadcasting stations from multiple prefectures alongside stations within Ishikawa Prefecture. Under normal circumstances, this is often understood merely as a long-distance reception phenomenon, but from a disaster perspective, it carries an entirely different meaning: even if stations within Ishikawa Prefecture were damaged, information from other prefectures could still be received. This dispersal of broadcasting stations across multiple directions contributes to improved redundancy, which is a critical advantage in geographically protruding regions such as the Noto Peninsula, where dependence on information supply from a single direction represents vulnerability. The reception results demonstrate that Suzu City is not merely peripheral to Ishikawa Prefecture's broadcasting network, but is electrically connected to broadcasting networks across the Sea of Japan coast — geographically a peninsula tip — but from the perspective of radio wave propagation, a junction of a wide-area broadcasting network. While the 2024 Noto Peninsula Earthquake produced physical isolation through road disruptions and communication failures, medium-wave broadcast waves propagate over the sea independently of such networks and therefore continue to function as an information transmission pathway during disasters without waiting for the restoration of communication and road infrastructure. The Suzu City reception survey thus demonstrates that medium-wave broadcasting forms not merely a regional broadcast but also a wide-area information network through maritime propagation, playing an important role in disaster-time information acquisition.

### 3.4 Significance of the High-Power NHK Radio 2 Broadcasting Network

Of particular note in this reception survey was the presence of the NHK Radio 2 broadcasting network, which was still operational at the time of the survey. NHK Radio 2 was constructed as a nationwide medium-wave network, with central stations comprising Sapporo (500 kW), Akita (500 kW), Shobu-Kuki (Kuki City, 500 kW), Osaka (300 kW), and Kumamoto (500 kW), the largest-power medium-wave broadcasting stations in Japan, with service areas incomparably wider than ordinary prefectural stations (Fig. 6). The Suzu City survey confirmed reception from multiple of these stations: good reception was obtained from Akita and Shobu-Kuki, with the Akita signal particularly benefiting from maritime propagation, and Osaka was also confirmed depending on the time of day, making clear that Suzu City can receive broadcast waves from multiple high-power stations. Importantly, these are not merely objects of long-distance reception, but can also function as disaster-time information infrastructure. Redundancy is critical for disaster countermeasures. As with power and communication systems, a broadcasting structure that is dependent on a single station is vulnerable. In contrast, the NHK Radio 2 network had high-power medium-wave broadcasting stations geographically distributed across multiple regions — Hokkaido, Tohoku, Kanto, Kansai, and Kyushu — which is an extremely favorable characteristic for disaster information reachability. Even if a station in one direction is damaged, information acquisition could continue through stations elsewhere, including outside the disaster-affected area.

At the time of the 2024 Noto Peninsula Earthquake, this nationwide network was still in operation, with high-power stations geographically distributed across Japan. This structure provided the potential for alternative information pathways if broadcasting facilities within Ishikawa Prefecture were damaged. In the September 2024 survey, reception in Suzu City was confirmed from the Akita and Tokyo stations, with Osaka also received depending on the time of day. Additional distant high-power stations, including Kumamoto and Sapporo, may also have been receivable under favorable nighttime ionospheric conditions through sky-wave propagation, although such reception was not confirmed in the present survey. In regions surrounded by the sea on three sides, such as the Noto Peninsula, medium-wave broadcast waves from high-power stations along the Sea of Japan and Pacific coast are relatively receivable. Notably, as shown in Section 3.3, Akita Ogata Broadcasting Station (375 km) showed a higher RSSI and lower noise than Shobu-Kuki Broadcasting Station (258 km), demonstrating that the proportion of maritime path segments — not transmission output or distance alone — plays an important role alongside transmission output and distance, and that stations such as Akita Ogata, located along the Sea of Japan coast, had particularly high potential for disaster-time information transmission on the Noto Peninsula. This underscores the importance of considering the propagation path characteristics alongside the output and distance when evaluating a nationwide medium-wave network. Furthermore, NHK Radio 2 operated on a nationwide uniform program schedule, providing a nationally consistent broadcasting service across regions.

The surveys and analyses in this study were based on conditions as they existed in September 2024, when NHK Radio 2 was in operation. It subsequently ended broadcasting on the night of March 29, 2026, drawing a curtain on 95 years of history against a backdrop of changes in its institutional role and usage conditions [11]. Nonetheless, the reception results obtained here demonstrate the potential value of such broadcasting networks during disasters, indicating that the significance of a nationwide high-power medium-wave network needs to be reassessed, particularly in peninsula and island regions where communications and transportation infrastructure tend to become vulnerable. The 2024 Noto Peninsula Earthquake provided an opportunity to empirically reconsider this value.

### 3.5 Disaster Information Reachability and Medium-Wave AM Broadcasting

The question from which this study originated is simple: "On that day, did information truly reach the people of Oku-Noto?" Tsunami warnings and evacuation information were disseminated through television, radio, the internet, and social media, and viewed solely from the perspective of information transmission, Japan's disaster information system can be said to have functioned. However, information transmission and reachability are not necessarily equivalent; television cannot be watched during a power outage, smartphone information becomes inaccessible if base stations go off-air, deployed satellite communication equipment is not necessarily easy for everyone to use, and even restored communication lines may be restricted by congestion and device usage environments. A large gap exists between the fact that "information was transmitted" and the fact that "residents were able to obtain information." This study proposes "Disaster Information Reachability" to evaluate this gap, defined as "the capacity to deliver necessary information to the people who need it at the very moment they need it." What matters in this definition is not the volume of information or communication speed: an advanced communication system is meaningless if a power outage prevents its use, a sophisticated smartphone application is meaningless if elderly residents cannot use it, and detailed evacuation information is meaningless if residents cannot receive it. What matters during a disaster is not technological sophistication but whether information ultimately reaches people — a critically important perspective given that the aging rate in Suzu City stands at approximately 52% and across all municipalities with a seismic intensity of upper-6 or above at approximately 44% [4]. Some elderly residents may not routinely use smartphones, and even regular users may have difficulty obtaining information during power outages and communication failures; such people can become “information-vulnerable,” defined here not by physical disability but by difficulty in information acquisition due to power outages, communication failures, aging, geographical isolation, and device usage environments — conditions that arose across a wide area in the 2024 Noto Peninsula Earthquake.

As the power outage data in Section 3.1 demonstrate, this gap is not merely theoretical; it was quantifiably documented in the fact that the majority of households in the highest-risk coastal areas had already lost power at the very moment the tsunami warning was issued [23, 24].

Meanwhile, the reception surveys in this study confirmed that medium-wave broadcasting stations from multiple regions — Niigata, Toyama, Akita, Tokyo, and Osaka — were receivable in Suzu City alongside stations within Ishikawa Prefecture, with AM broadcasting continuously receivable throughout the long-distance drive from Kanazawa to Suzu City, with almost none of the frequent rescanning or unreceivable zones observed with FM. This carries critically important meaning beyond mere radio wave propagation: even if one broadcasting station goes off-air, waves from other directions remain available, and even if one information pathway is lost, information arrives through another; that is, medium-wave AM broadcasting possesses high redundancy as an information infrastructure. The contrast with the FM reception results is highly instructive: all FM stations except the nearby Noto Town relay station (11 km) showed RSSI values of only 9–17, with audible reception extremely difficult, including the FM key stations in Nonoichi City (116 km) within Ishikawa Prefecture itself, where the mountainous terrain of the Noto Peninsula obstructs the VHF line-of-sight propagation. Because FM broadcasting depends heavily on nearby relay stations — which the 2024 earthquake's road disruptions caused to go off-air in some cases — reception quality deteriorates markedly when such a single point of failure is lost, if the Noto Town relay station were to go off-air, only the distant Nonoichi station could be relied upon, making reception extremely difficult. AM broadcasting, in contrast, enables direct reception from high-power key stations through maritime propagation from multiple directions, without dependence on relay stations — a fundamental difference in redundancy with important implications for the design of disaster-time information infrastructure in peninsular regions.

Redundancy is an extremely important concept in disaster countermeasures: roads, communication lines, and power supplies benefit from multiple pathways rather than one, and information transmission is no exception. Medium-wave broadcast waves from various locations along the Sea of Japan coast confirmed in Suzu City demonstrate that the Noto Peninsula can receive information from multiple directions through maritime propagation. Beyond transmission, AM receivers are also extremely simple, operable on batteries for extended periods, and in some cases, receivable via passive radio. The HOOPRA developed by the author confirmed reception without external power or batteries, demonstrating that broadcast waves can serve as an energy source and information medium. In this sense, the HOOPRA results should not be interpreted merely as a receiver demonstration, but as an example of how medium-wave broadcast waves can provide both information and usable electromagnetic energy, even when ordinary power and communication infrastructure are unavailable. The reception confirmations shown in Fig. 7 illustrate this practical aspect. By magnetically coupling the HOOPRA to a commercial receiver, weak medium-wave broadcasts become more readily identifiable without requiring any external power source. This supports the broader argument that medium-wave AM broadcasting has disaster-time value not only as a transmission system but also as an information environment that can be accessed with simple, low-power, or even passive receiving devices. Unlike communication systems, broadcasting suffers no performance degradation from congestion regardless of how many users receive it simultaneously — a "simultaneous broadcast capability" that is a significant advantage when, as in the 2024 Noto Peninsula Earthquake, large numbers of residents simultaneously require evacuation, tsunami, road, and water supply information.

The NHK Radio 2 network examined in this study, which formed a nationwide medium-wave network through high-power stations in Sapporo, Akita, Tokyo, Osaka, and Kumamoto, can be interpreted from this perspective as a geographically distributed information redundancy system deployed across the country. At the time of the earthquake, this network was still in operation. The September 2024 survey subsequently confirmed reception from multiple high-power stations in Suzu City, demonstrating the potential of such a network as an information reachability pathway during disasters. The wide-area reception capability through maritime propagation confirmed here is an example where the effect appeared particularly prominent given the Noto Peninsula's geography, but its significance is not limited to this region. Japan has numerous peninsula and island regions, and in anticipated future disasters, including the Nankai Trough megaquake, Tokyo Inland Earthquake, Japan Trench and Kuril Trench earthquake, and Mount Fuji eruption [6], power outages, communication failures, road disruptions, and information isolation are fully anticipated. Securing disaster information reachability in such situations requires maintaining diverse means of information transmission. The results of this study demonstrate that medium-wave AM broadcasting functions not as a relic of the past but as an important social infrastructure supporting it — a component of social infrastructure alongside roads, electricity, and communications, whose value becomes more apparent during disasters than under normal conditions.

The author considers that evaluation of disaster information infrastructure has tended to prioritize "how much information can be transmitted" and "how advanced a communication technology can be deployed," when what truly matters is "whether it delivers to the end." During disasters, simple and robust systems can function more effectively than cutting-edge communication technologies; medium-wave AM broadcasting is a prime example. In response to large-scale disaster countermeasures following the 2024 Noto Peninsula Earthquake, the Ministry of Internal Affairs and Communications identified the strengthening of mobile phone base station resilience, fiber-optic conversion and route diversification of cable television networks, and intercarrier roaming as challenges [22]. While that document records specific outage cases and restoration processes for FM relay station damage in detail, there is no mention of medium-wave AM broadcasting [22]. This absence does not establish its reception performance during the disaster, but indicates that the disaster-time role and reachability of medium-wave AM broadcasting were not explicitly addressed in that policy discussion. The results of this study demonstrate that the maintenance of multiple information transmission means, including medium-wave AM broadcasting, should be reconsidered in the design of future disaster information infrastructure. Medium-wave AM broadcasting is an information infrastructure that "delivers to the end," deserving reassessment as important social infrastructure supporting disaster information reachability — a question the 2024 Noto Peninsula Earthquake poses once again to society.

**4. Conclusion**

This study conducted a medium-wave broadcast reception survey in Suzu City, Ishikawa Prefecture, using the 2024 Noto Peninsula Earthquake as a case study, examining the role of medium-wave AM broadcasting from the perspective of Disaster Information Reachability. The results confirmed that medium-wave broadcasting stations from distant prefectures — Niigata, Toyama, Akita, Tokyo, and Osaka — were receivable in Suzu City alongside stations within Ishikawa Prefecture, and that AM broadcasting was stably receivable across the wide-area mobile survey route from Kanazawa City to Suzu City, in contrast to FM broadcasting, which showed large fluctuations, unreceivable zones, and frequent rescanning owing to terrain effects. These results demonstrate that maritime propagation paths over the Sea of Japan support the wide-area reception of medium-wave broadcasting. As described above, the 2024 Noto Peninsula Earthquake caused widespread power outages, communication failures, mobile phone base station outages, fixed telephone network failures, television relay station outages, and cable television network damage, with telephone interviews confirming difficulties in using communication lines, inability to watch television, and operational challenges with satellite communication equipment. These characteristics indicate that medium-wave AM broadcasting can serve as a means of information acquisition under such disaster conditions through its wide-area simultaneous broadcast capability, low-power reception, maritime propagation characteristics, and high redundancy. The fact that stations from multiple prefectures were receivable in Suzu City means that information acquisition pathways do not depend on a single broadcasting station or communication means, which is an extremely important characteristic of disaster-time information infrastructure. This study's central premise — that what matters in disaster countermeasures is not information transmission itself, but the actual ability of disaster victims to obtain information — is reinforced by the fact that while FM relay station damage was documented in detail in policy discussions following the earthquake, there is no mention of medium-wave AM broadcasting [22], demonstrating the necessity of a shift from information-transmission-centered to information-reachability-centered discussion.

In recent years, the center of information transmission has shifted to the Internet and smartphones, but as the 2024 Noto Peninsula Earthquake clarified, communication infrastructure itself can be damaged and rendered unusable in large-scale disasters, and excessive dependence on it leads to vulnerability, particularly in peninsula and island regions where aging and depopulation are advancing rapidly [4]. Medium-wave AM broadcasting, in contrast, possesses wide-area reachability from multiple directions and high redundancy, suffers no performance degradation from congestion or increased users, and uses inexpensive, low-power receivers capable of extended battery operation and passive reception — characteristics that make it extremely effective for large-scale disasters. The NHK Radio 2 network, from which multiple high-power stations were received in this study and which formed a nationwide medium-wave network through high-power stations in Sapporo, Akita, Tokyo, Osaka, and Kumamoto, was still in operation at the time of the earthquake and was subsequently closed on the night of March 29, 2026 [11]. Nonetheless, the reception results obtained in this study demonstrate the potential value of such a network during disasters and the importance of nationwide high-power medium-wave networks in securing information redundancy.

This study, through the case of the 2024 Noto Peninsula Earthquake, has shown the importance of disaster information reachability in peninsula disaster prevention and has provided foundational knowledge for reassessing the social value of medium-wave AM broadcasting. The concept of Disaster Information Reachability proposed here can serve as an evaluation indicator for future disaster information systems. Future work will extend similar reception surveys to the peninsula and island regions beyond Noto; pursue quantitative evaluation methods integrating receivable station counts, reception quality, broadcasting station and population distribution, and disaster risk information; and examine the design of multiple-layered information infrastructure combining broadcasting and communications for anticipated large-scale disasters, including the Nankai Trough megaquake, Tokyo Inland Earthquake, Japan Trench and Kuril Trench earthquake, and Mount Fuji eruption [6]. What matters during disasters is not "what information was transmitted" but "whether that information truly reached people who needed it." Medium-wave AM broadcasting is an information infrastructure that can serve as a compelling answer to this question.

**Acknowledgments**

The author thanks Ms. Naoko Shoji for assistance with equipment transport, on-site measurements, and valuable discussions. This work was supported by JSPS KAKENHI (grant number JP24K15388).

**Conflict of Interest**

The authors declare no conflicts of interest related to this study.

**Data Availability Statement**

Due to the nature of the field measurements and the experimental environment, the data are not publicly available but may be available from the corresponding author upon reasonable request.

**Legend of Figures**

Figure 1. Geographical overview of the Noto Peninsula and survey area. A map of the Noto Peninsula is presented. The survey routes along the Noto Satoyama Expressway (Kanazawa City to Anamizu Town) and Suzu Road (Anamizu Town to Suzu City), the location of Suzu City Hall (survey point), and the locations of Wajima City, Anamizu Town, Nanao City, and Kanazawa City are indicated. The figure illustrates the terrain of the Noto Peninsula, which is surrounded on three sides by the Sea of Japan and Toyama Bay.

Figure 2. Distribution of the 23 peninsula development promotion regions in Japan. Based on the map of Peninsula Development Promotion Implementation Areas published by the Ministry of Land, Infrastructure, Transport, and Tourism [3], this figure summarizes the distribution of the 23 designated peninsula development promotion regions in Japan. The Noto region, corresponding to the Noto Peninsula, is indicated by a larger bold text as the case study area.

Figure 3. Survey route and survey conditions at Suzu City Hall. (A) Driving conditions on the Noto Satoyama Expressway near the entrance from Kanazawa; (B) driving conditions near a repaired section of the expressway where the road had collapsed; (C) measurement experiment in the parking lot of Suzu City Hall; (D) reception-enhancement experiment using the 80 cm HOOPRA as an external antenna, corresponding to conditions B, D, and F of the comparison experiment shown in Fig. 7 and described in Section 3.5. The on-site environment on the day of the survey (September 15, 2024) is shown.

Figure 4. Layout of medium-wave broadcasting stations around Suzu City and received signal strength. The broadcasting station positions and distances to Suzu City of medium-wave broadcasting stations, whose reception was confirmed in the area surrounding Suzu City Hall, are shown. Propagation paths from the broadcasting stations within Ishikawa Prefecture (Nonoichi City, Nanao City, and Wajima City), Toyama Prefecture (Toyama City), and Niigata Prefecture (Niigata City, Kashiwazaki City, Joetsu City, and Itoigawa City) to Suzu City are indicated by arrows. The values in the figure represent the RSSI (dBμ) measured at the Suzu City Hall survey point for signals from each broadcasting station.

Figure 5. Reception of distant medium-wave broadcasting stations through maritime propagation. The broadcasting station positions and propagation paths of distant high-power stations, whose reception was confirmed in Suzu City, are shown. The propagation path from the NHK Akita Radio 2 (JOUB, 774 kHz, 500 kW, Akita Ogata Broadcasting Station, RSSI 40) is almost entirely maritime propagation over the Sea of Japan, whereas the propagation path from the NHK Tokyo Radio 1 and Radio 2 (Shobu-Kuki Broadcasting Station, Kuki City, RSSI 35–37) includes many overland propagation segments that pass through the Kanto Plain and mountainous areas. The differences in RSSI and noise levels between the two indicate that the proportion of maritime segments in the propagation path is an important factor influencing reception quality.

Figure 6. Nationwide layout of NHK Radio 2 high-power broadcasting stations. The geographical layout of the major broadcasting stations of NHK Radio 2 is shown. The positions of Sapporo Broadcasting Station (JOIB, 747 kHz, 500 kW), Akita Ogata Broadcasting Station (JOUB, 774 kHz, 500 kW), Shobu-Kuki Broadcasting Station (JOAB, 693 kHz, 500 kW, Kuki City), Osaka Broadcasting Station (JOBB, 828 kHz, 300 kW), and Kumamoto Broadcasting Station (JOGB, 873 kHz, 500 kW) are indicated, with the circle size representing the transmission output scale. The survey was conducted in September 2024, when NHK Radio 2 broadcasting was still in operation.

Figure 7. Comparison of reception with and without HOOPRA assistance at Suzu City Hall. (A) NHK Kanazawa Radio 2 (JOJB, 1386 kHz, 10 kW) without HOOPRA, RSSI 15; (B) NHK Kanazawa Radio 2 with HOOPRA assistance, RSSI 46; (C) NHK Kanazawa Radio 1 (JOJK, 1224 kHz, 10 kW) without HOOPRA, RSSI 24; (D) NHK Kanazawa Radio 1 with HOOPRA assistance, RSSI 55; (E) Fukui Broadcasting (JOPR, 864 kHz, 5 kW) without HOOPRA, RSSI 24; and (F) Fukui Broadcasting with HOOPRA assistance, RSSI 50. Under all conditions, the ferrite bar antenna of the receiver and loop plane of the HOOPRA were oriented toward the direction of arrival of the broadcast wave. The RSSI values are also visible on the receiver display in each photograph.

**Legend of Tables**

Table 1. Results of medium-wave AM broadcast reception survey in Suzu City

Fixed-point reception survey results for medium-wave AM broadcasting conducted in the area surrounding Suzu City Hall on September 15, 2024, are presented. The broadcasting station name, call sign, broadcasting station location, frequency (kHz), transmission output (kW), RSSI (dBμ), and straight-line distance from the broadcasting station to Suzu City Hall (km) are shown. "—" indicates stations for which reception was not confirmed. "Relay station (off-air at time of survey)" in the comment column indicates suspension of operation due to the effects of the 2024 Noto Peninsula Earthquake.

Table 2. Results of FM broadcast reception survey in Suzu City

Fixed-point reception survey results for FM broadcasting conducted in the area surrounding Suzu City Hall on September 15, 2024, are presented. The broadcasting station name, call sign, broadcasting station location, frequency (MHz), transmission output (kW), RSSI (dBμ), and straight-line distance from the broadcasting station to Suzu City Hall (km) are shown. FM broadcast reception used only the built-in rod antenna of the commercially available radio (PL-368). In the comment column, “Relay Broadcasting Station; reception confirmed” indicates stable reception from a relay broadcasting station, “unstable reception” indicates RSSI values of 11 to 20, and “difficult reception” indicates RSSI values of 10 or below. “Supplementary broadcast” refers to FM supplementary broadcasting (FM complementary broadcasting) operated by AM broadcasting stations. The “Suzu relay stations,” despite their name, are located in Aza Akeno, Hosu-gun, Noto Town, Ishikawa Prefecture, in the vicinity of Takakura-yama, approximately 11 km from the Suzu City Hall survey point.

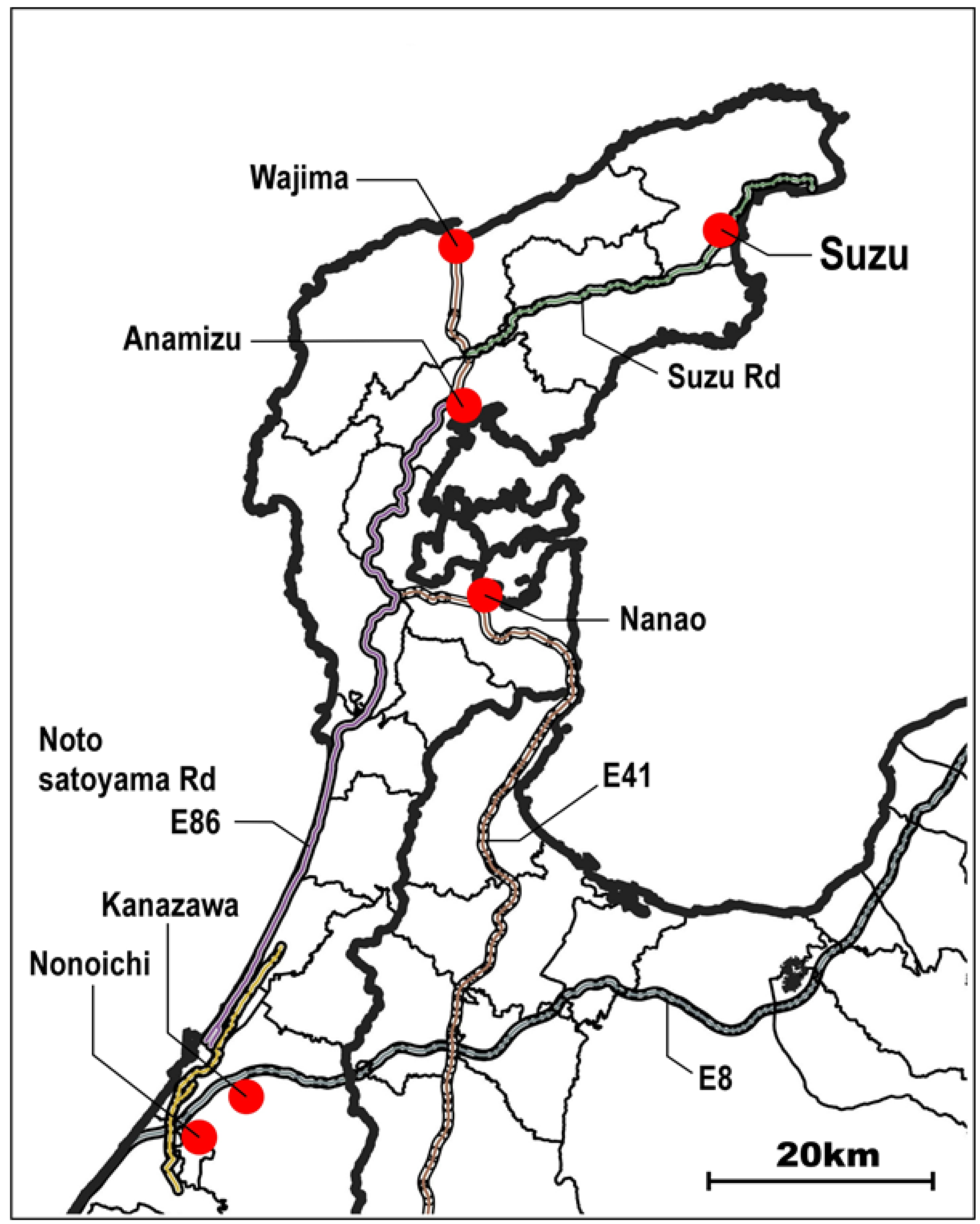

Figure 1

E. Shoji

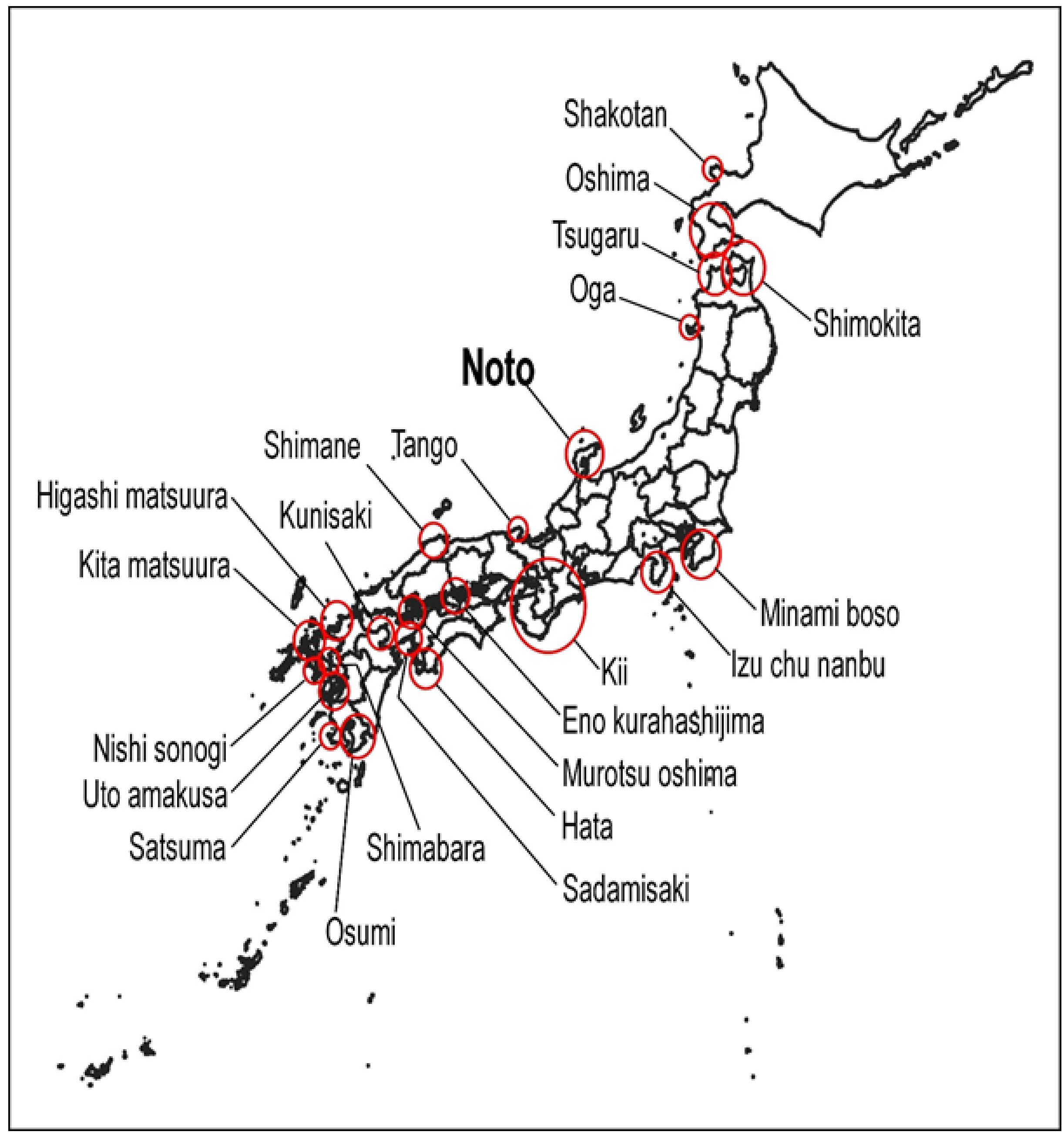

Figure 2
E. Shoji

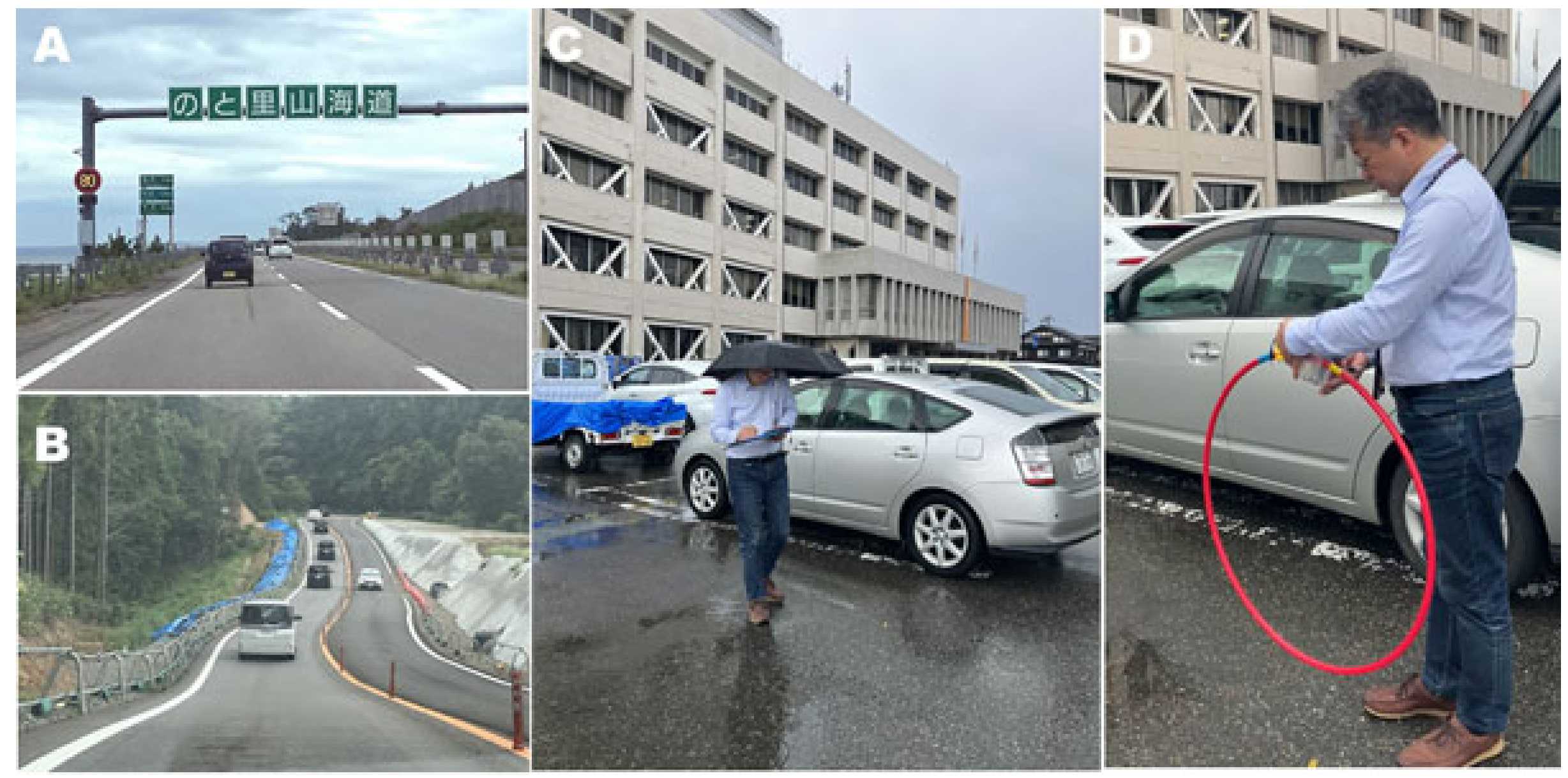

Figure 3

E. Shoji

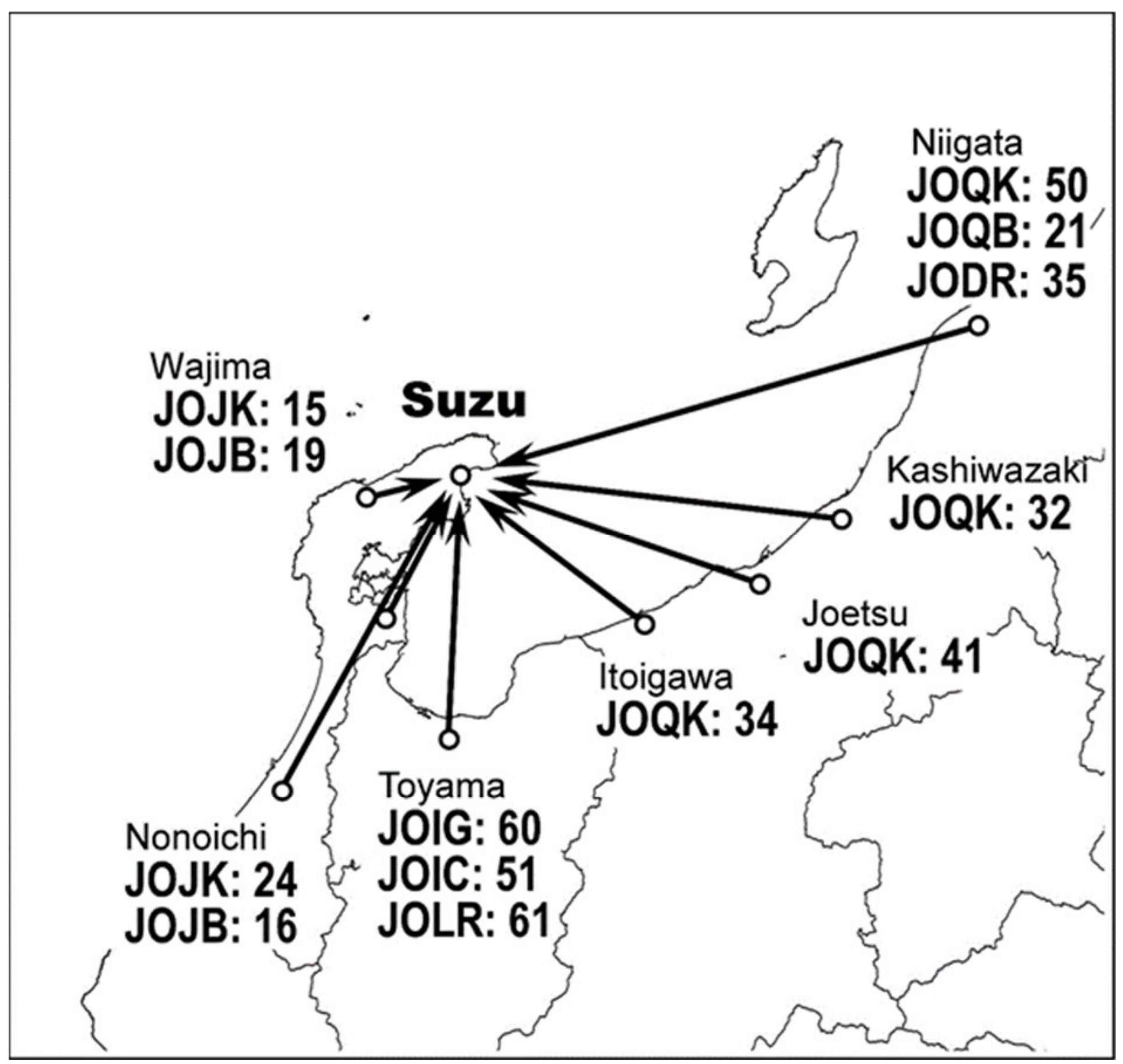

Figure 4
E. Shoji

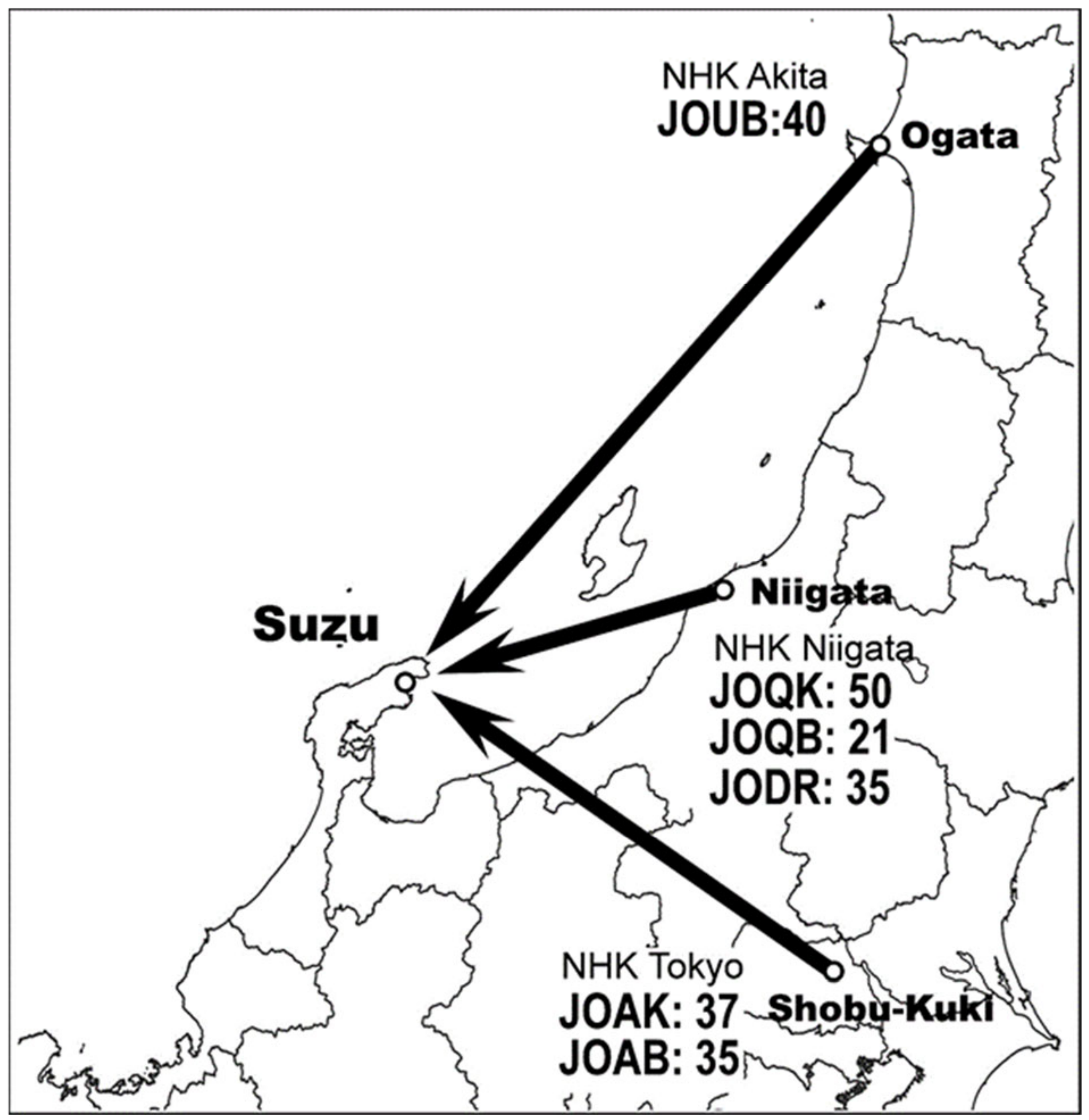

Figure 5
E. Shoji

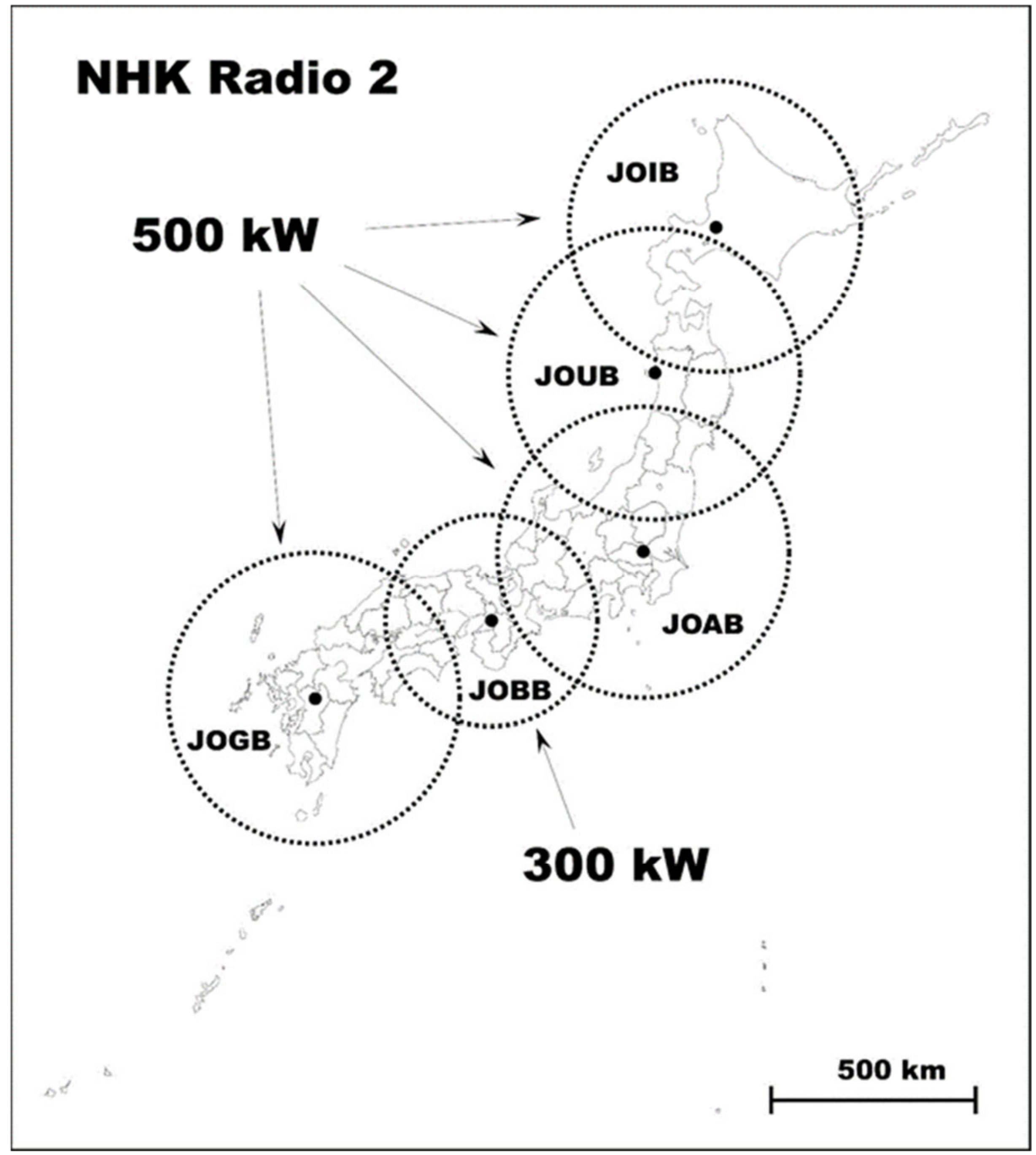

Figure 6
E. Shoji

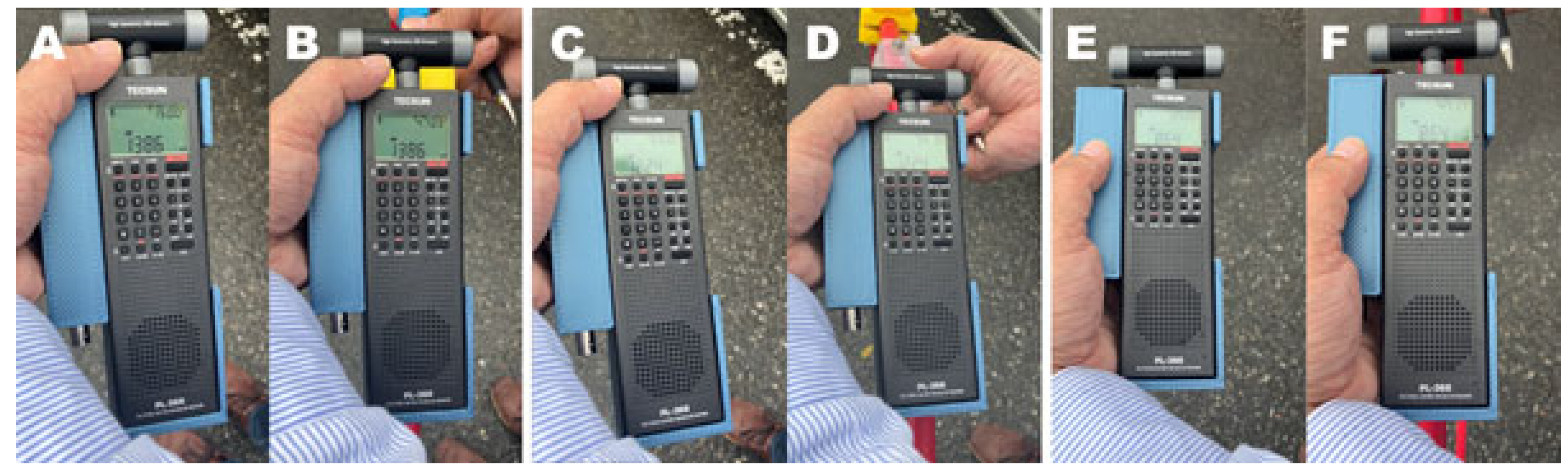

Figure 7
E. Shoji

Table 1 AM Broadcasting Reception Survey Results in Suzu City.

| Station | Call Sign | Broadcasting Station | Frequency (kHz) | Power (kW) | RSSI (dBμ) | Distance (km) | Note |
|---|---|---|---|---|---|---|---|
| **Niigata** | | | | | | | |
| NHK Radio 1 | JOQK | Niigata City | 837 | 10 | 50 | 145 | |
| NHK Radio 2 | JOQB | Niigata City | 1593 | 10 | 21 | 145 | |
| NHK Radio 1 | JOQK | Itoigawa City | 999 | 0.1 | 34 | 68 | Relay Broadcasting Station |
| NHK Radio 1 | JOQK | Joetsu City | 792 | 1 | 41 | 91 | Relay Broadcasting Station |
| NHK Radio 1 | JOQK | Kashiwazaki City | 981 | 0.1 | 32 | 114 | Relay Broadcasting Station |
| Niigata Broadcasting (BSN) | JODR | Niigata City | 1116 | 5 | 35 | 168 | |
| **Toyama** | | | | | | | |
| NHK Radio 1 | JOIG | Toyama City | 648 | 5 | 60 | 81 | |
| NHK Radio 2 | JOIC | Toyama City | 1035 | 1 | 51 | 81 | |
| Kitanippon Broadcasting (KNB) | JOLR | Toyama City | 738 | 5 | 61 | 81 | |
| **Ishikawa** | | | | | | | |
| NHK Radio 1 | JOJK | Nonoichi City | 1224 | 10 | 24 | 116 | |
| NHK Radio 2 | JOJB | Nonoichi City | 1386 | 10 | 15 | 116 | |
| NHK Radio 1 | JOJK | Nanao City | 540 | 1 | 46 | 53 | Relay Broadcasting Station |
| NHK Radio 2 | JOJB | Nanao City | 1467 | 0.1 | 9 | 53 | Relay Broadcasting Station |
| NHK Radio 1 | JOJK | Wajima City | 1584 | 0.1 | 15 | 32 | Relay Broadcasting Station |
| NHK Radio 2 | JOJB | Wajima City | 1359 | 0.1 | 19 | 32 | Relay Broadcasting Station |
| Hokuriku Broadcasting (MRO) | JOMR | Nonoichi City | 1107 | 5 | 24 | 116 | |
| Hokuriku Broadcasting (MRO) | JOMR | Nanao City | 1107 | 1 | — | 53 | Relay Broadcasting Station; off-air during survey |
| Hokuriku Broadcasting (MRO) | JOMR | Wajima City | 1107 | 0.1 | — | 32 | Relay Broadcasting Station; off-air during survey |
| **Fukui** | | | | | | | |
| NHK Radio 1 | JOFG | Fukui City | 927 | 5 | 18 | 179 | |
| NHK Radio 2 | JOFC | Fukui City | 1521 | 1 | — | 179 | difficult reception |
| Fukui Broadcasting (FBC) | JOPR | Sakai City | 864 | 5 | 24 | 167 | |
| **Akita** | | | | | | | |
| NHK Radio 2 | JOUB | Ogata Village | 774 | 500 | 40 | 375 | |
| **Tokyo** | | | | | | | |
| NHK Radio 1 | JOAK | Kuki City | 594 | 300 | 37 | 258 | |
| NHK Radio 2 | JOAB | Kuki City | 693 | 500 | 35 | 258 | |

Table 1

E. Shoji

Table 2 FM Broadcasting Reception Survey Results in Suzu City.

| Station | Call Sign | Broadcasting Station | Frequency (MHz) | Power (kW) | RSSI (dBμ) | Distance (km) | Note |
|---|---|---|---|---|---|---|---|
| **Niigata** | | | | | | | |
| NHK-FM | JOQK-FM | Niigata City | 82.3 | 1 | 17 | 142 | Unstable reception |
| FM Niigata (FM-NIIGATA) | JOXU-FM | Niigata City | 77.5 | 1 | 17 | 142 | Unstable reception |
| Niigata Broadcasting (BSN) | Supplementary broadcast | Niigata City | 92.7 | 1 | 14 | 142 | Unstable reception |
| **Toyama** | | | | | | | |
| NHK-FM | JOIG-FM | Toyama City | 81.5 | 1 | 14 | 81 | Unstable reception |
| Toyama FM Broadcasting (FMT) | JOOU-FM | Toyama City | 82.7 | 1 | 17 | 81 | Unstable reception |
| Kitanippon Broadcasting (KNB) | Supplementary broadcast | Toyama City | 90.2 | 1 | 16 | 81 | Unstable reception |
| **Ishikawa** | | | | | | | |
| NHK-FM | JOJK-FM | Nonoichi City | 82.2 | 1 | 16 | 116 | Unstable reception |
| NHK-FM | JOJK-FM | Nanao City | 84.4 | 0.1 | 14 | 53 | Unstable reception |
| NHK-FM | JOJK-FM | Noto Town | 83.2 | 0.1 | 37 | 11 | Relay Broadcasting Station; reception confirmed |
| FM Ishikawa (Hello Five) | JOHV-FM | Nonoichi City | 80.5 | 1 | 12 | 116 | Unstable reception |
| FM Ishikawa (Hello Five) | JOHV-FM | Nanao City | 78.4 | 0.1 | 14 | 53 | Unstable reception |
| FM Ishikawa (Hello Five) | JOHV-FM | Noto Town | 81.9 | 0.1 | 34 | 11 | Relay Broadcasting Station; reception confirmed |
| Hokuriku Broadcasting (MRO) | Supplementary broadcast | Nonoichi City | 94 | 1 | 10 | 116 | difficult reception |
| Hokuriku Broadcasting (MRO) | Supplementary broadcast | Nanao City | 88.6 | 0.1 | 13 | 53 | Unstable reception |
| Hokuriku Broadcasting (MRO) | Supplementary broadcast | Noto Town | 76.7 | 0.1 | 35 | 11 | Relay Broadcasting Station; reception confirmed |
| **Fukui** | | | | | | | |
| NHK-FM | JOFG-FM | Fukui City | 83.4 | 1 | 12 | 177 | Unstable reception |
| FM Fukui (FM Fukui) | JOLU-FM | Fukui City | 76.1 | 1 | 17 | 177 | Unstable reception |
| Fukui Broadcasting (FBC) | Supplementary broadcast | Fukui City | 94.6 | 1 | 9 | 177 | difficult reception |

Table 2
E. Shoji